\documentclass[12pt]{article}
\usepackage{amssymb}
\usepackage{epsfig}

\catcode`\@=11
\def\numberbysection{\@addtoreset{equation}{section}
        \def\theequation{\thesection.\arabic{equation}}}

\def\beq{\begin{equation}}
\def\eeq{\end{equation}}
\numberbysection
\begin{document}
\begin{titlepage}
\begin{center}
\hfill DFF  1/11/05 \\
\vskip 1.in {\Large \bf $U(2)$ projectors and 't Hooft-Polyakov
monopoles on a fuzzy sphere} \vskip 0.5in P. Valtancoli
\\[.2in]
{\em Dipartimento di Fisica, Polo Scientifico Universit\'a di Firenze \\
and INFN, Sezione di Firenze (Italy)\\
Via G. Sansone 1, 50019 Sesto Fiorentino, Italy}
\end{center}
\vskip .5in
\begin{abstract}
We show how to generalize our method, based on projective modules
and matrix models, which enabled us to derive noncommutative
monopoles on a fuzzy sphere, to the non-abelian case, recovering
known results in literature. We then discuss a possible candidate
for deforming the commutative Chern class to the non-commutative
case.
\end{abstract}
\medskip
\end{titlepage}
\pagenumbering{arabic}
\section{ Introduction }

Recently the study of topologically nontrivial configurations
\cite{3}-\cite{4}-\cite{5} on a noncommutative manifold \cite{1} has
received growing interest. In this framework we have set up a
general method with which all the noncommutative topological
configurations for a $ U(1) $ gauge group could be studied
exhaustively \cite{23}.

Basic elements of such a method are a combination of mathematical
and physical concepts. To identify easily the nontrivial topological
configurations we have made use of the concept of projector
\cite{6}-\cite{7}-\cite{21}, obtaining their classification on a
fuzzy sphere for a $U(1)$ gauge group \cite{2}.

With a subsequent study we have introduced a natural way for
defining noncommutative actions on a fuzzy sphere
\cite{8}-\cite{9}-\cite{10}-\cite{11}-\cite{12}-\cite{13}, by using
the matrix model approach
\cite{14}-\cite{15}-\cite{16}-\cite{17}-\cite{18}-\cite{19}-\cite{20}.
Finally with a simple link between projectors and matrix models we
have associated to the noncommutative projectors the corresponding
noncommutative connections \cite{23}-\cite{22}-\cite{24},
generalizing Dirac monopoles on a fuzzy sphere \cite{23}-\cite{25}.

To demonstrate that our method is quite general, we have studied in
this work the case of a nonabelian gauge group, and , as we will
show next, we are able to obtain the same results known in
literature \cite{25}.

Starting from a noncommutative projector with entries belonging to $
U(2) $, we show that it is possible to reconstruct, along the same
lines of the abelian case, a solution of the matrix model equations
of motion, that coincides, in the classical limit, with the 't
Hooft-Polyakov monopoles for $SU(2)$ gauge group.

In the final part of this work we discuss a possible candidate for
deforming the Chern class and its corresponding topological number.
The action we propose is the $3d$ Chern Simons action, which enjoys
the property of invariance under deformed diffeomorphisms \cite{26}.
While the nonabelian case works fine without any problem, the
abelian one appears more obscure and requires a deeper analysis of
the commutative limit, to reach smoothness with the Chern class.

\section{Review of abelian noncommutative monopoles}

The aim of this work is to describe how abelian and non-abelian
monopoles can be deformed on a non-commutative sphere. Our research
tries to cover both mathematical and physical issues, filling the
gap between the projective module point of view and the matrix model
formalism, which incorporates the natural definition of a
non-commutative gauge theory on a non-commutative manifold.

Since the abelian case has been already worked out in ref.
\cite{23}, we briefly remind the principal features in this section
so that the reader can easily identify common features and
differences with the corresponding non-abelian case.

The starting point is the definition of the fuzzy sphere algebra

\begin{eqnarray} [ \hat{x}_i , \hat{x}_j ] & = & i \alpha \epsilon_{ijk} \hat{x}_k
\ \ \ \ \ \sum_i { ( \hat{x}_i )}^2 = R^2  \nonumber \\
\alpha & = & \frac{2 R}{\sqrt{ N ( N+2 )} } \label{21}
\end{eqnarray}

which produces an useful finite truncation of the
infinite-dimensional function space of the classical sphere.

Such an algebra can be simply obtained by quantizing the Hopf
fibration $S^3 \rightarrow S^2$, which is defined in terms of two
complex coordinates, constrained to the $S^3$ sphere :

\begin{eqnarray}
x_1 & = & z_0 \overline{z}_1 + z_1 \overline{z}_0 \nonumber \\
x_2 & = & i ( z_0 \overline{z}_1 - z_1 \overline{z}_0 ) \nonumber \\
x_3 & = & | z_0  |^2 - | z_1 |^2 \ \ \ \ \ {  | z_0 | }^2 + { | z_1
| }^2 = 1 \label{22} \end{eqnarray}

Substituting into this mapping (\ref{22}) the complex coordinates
with oscillator operators $a_i$ satisfying the algebra

\beq [ a_i , a_j^{\dagger} ] = \delta_{ij} \label{23} \eeq

the corresponding real combinations are representations of the fuzzy
sphere algebra :

\begin{eqnarray}
\hat{x}_1 & = & \frac{\hat{\alpha}}{2} ( a_0 a_1^{\dagger} + a_1
a_0^{\dagger} )  \nonumber \\
\hat{x}_2 & = & i \frac{\hat{\alpha}}{2} ( a_0 a_1^{\dagger} - a_1
a_0^{\dagger} ) \nonumber \\
\hat{x}_3 & = & \frac{\hat{\alpha}}{2} ( a_0 a_0^{\dagger} - a_1
a_1^{\dagger} ) \nonumber \\
\hat{N} & = & a^\dagger_0 a_0 + a^\dagger_1 a_1 \label{24}
\end{eqnarray}

The $\hat{\alpha}$ operator has to be defined yet. Restricting the
action of the oscillators on representations with fixed total number
$ \hat{N} = N $, the $\hat{\alpha}$ operator can be taken as the
$\alpha $ constant

\beq \hat{\alpha} \rightarrow \alpha = \frac{ 2 R }{ \sqrt{ N ( N+2
)} } \label{25} \eeq

This relation is particularly useful to generalize the monopole
projectors at a non-commutative level. Let us recall the classical
$k=1$ case, i.e. the simplest monopole configuration, which is
defined by the following projector $P_1$, function of the complex
coordinates $z_i$:

\begin{eqnarray}
P_1 \ \ & = & | \psi_1 >< \psi_1 | \nonumber \\
| \psi_1 > & = &  \left( \begin{array}{c} z_0 \\ z_1 \end{array}
\right) \label{26} \end{eqnarray}

The normalization condition of this vector

\beq < \psi_1 | \psi_1 > = {| z_0 |}^2 + {| z_1 |}^2 = 1 \label{27}
\eeq

is satisfied, since the complex coordinates belong to $S^3$, as in
formula (\ref{22}).

In our first paper \cite{2} we noticed that the natural
non-commutative extension, based on quantizing the Hopf fibration (
eq. (\ref{23}) )

\begin{eqnarray}
P_1 \ \  & = & | \psi_1 > < \psi_1 | \nonumber \\
| \psi_1 > &  = & N_1 \left( \begin{array}{c} a_0 \\ a_1 \end{array}
\right) \label{28}
 \end{eqnarray}

really works by choosing the normalization factor as

\begin{eqnarray} < \psi_1 | \psi_1 > & =  & 1  \nonumber \\
N_1 =  N_1 ( \hat{N} ) & = & \frac{1}{\sqrt{ \hat{N} + 1 }}
\label{29} \end{eqnarray}

Since the fuzzy sphere algebra is a finite type non-commutative
geometry, the trace of the non-commutative projector is an integer
number :

\beq Tr P_1 = Tr | \psi_1 > < \psi_1 | =  N+2  < Tr 1_P = 2 ( N+1 )
\label{210} \eeq

It is possible to repeat the same procedure for monopoles with
negative charge. For example the projector corresponding to $ k = -1
$ is defined as :

\begin{eqnarray}
P_{-1} \ \  & = & | \psi_{-1} > < \psi_{-1} | \nonumber \\
| \psi_{-1} > &  = & N_{-1} \left( \begin{array}{c} a^{\dagger}_0 \\
a^{\dagger}_1
\end{array} \right) \label{211}
 \end{eqnarray}

where the normalization factor is determined by the condition

\beq < \psi_{-1} | \psi_{-1} > = 1 \ \ \ \  \Rightarrow N_{-1} =
N_{-1} ( \hat{N} ) = \frac{1}{\sqrt{ \hat{N} +1 }} \label{212} \eeq

Again the trace of $P_{-1}$ is an integer number:

\beq Tr P_{-1} = N < Tr 1_P = 2 ( N+1 ) \label{213} \eeq

Proving that these analogues of the topological excitations of
Yang-Mills theory on a sphere satisfy the non-commutative equations
of motion  requires the definition of a Yang-Mills theory on a fuzzy
sphere.

This proof has been firstly achieved in ref. \cite{23} with the
introduction of a matrix model $X_i ( i = 1, .., 3 )$, whose
equations of motion contain the fuzzy sphere algebra as particular
solutions:

\beq S( \lambda ) = S_0 + \lambda S_1 =
 - \frac{1}{g^2} Tr [
\frac{1}{4} [X_i,X_j][X_i,X_j] - \frac{2}{3}i\lambda \alpha
\epsilon^{ijk} X_i X_j X_k + \alpha^2 ( 1-\lambda ) X_i X_i ]
\label{214} \eeq

Both actions $S_0$ and $S_1$ separately contain the fuzzy sphere
algebra as a particular solution, and $\lambda$ is a generic
parameter.

The usual gauge connection is obtained by separating the generic
matrix $X_i$ from the background $L_i$ ( in the following we will
change the variable $X_i $ into $ \alpha X_i $ ) :

\beq X_i = L_i + A_i \ \ \ \ [ L_i , . ] \rightarrow_{\alpha
\rightarrow 0} - i k_i^a \partial_a \label{215} \eeq

The background $ L_i $ is equal to the classical Lie derivative on
the sphere in the $\alpha \rightarrow 0 $ ( classical limit ).
Instead the fluctuation $A_i$ is a linear combination of the
Yang-Mills connection $A_a ( \Omega ) $ and an auxiliary scalar
field ( in the adjoint representation ) $ \phi $:

\beq A_i ( \Omega ) = k_i^a A_a ( \Omega ) + \frac{x_i}{R} \phi (
\Omega ) \ \ \ \ \Omega = ( \theta, \phi ) \label{216} \eeq

A gauge covariant field strength can be defined as

\begin{eqnarray}
F_{ij} & = & [ X_i , X_j ] - i \epsilon_{ijk} X_k = \nonumber \\
& = & [ L_i , A_j ] - [ L_j , A_i ] + [ A_i , A_j ] - i
\epsilon_{ijk} A_k \label{217} \end{eqnarray}

In the classical limit it is useful to calculate it in terms of the
component fields

\beq F_{ij} ( \Omega ) = k_i^a k_j^b F_{ab} + \frac{i}{R}
\epsilon_{ijk} x_k \phi - \frac{x_i}{R} k_j^a D_a \phi +
\frac{x_j}{R} k_i^a D_a \phi \label{218} \eeq

where

\begin{eqnarray}
F_{ab} & = & -i ( \partial_a A_b - \partial_b A_a ) + [ A_a, A_b ]
\nonumber \\
D_a \phi & = & - i \partial_a \phi + [ A_a , \phi ] \label{219}
 \end{eqnarray}

By inserting these formulas into the action $S(\lambda)$, its value
in the classical limit is the general action:

\begin{eqnarray}
S( \lambda ) & = & - \frac{1}{4 g^2_{YM} } \int d \Omega [ ( F_{ab}
+ ( 4 - 2 \lambda ) i \epsilon_{ab} \phi \sqrt{g} )( F^{ab} + ( 4 -
2 \lambda ) i \epsilon^{ab} \frac{\phi }{\sqrt{g}} ) +
\nonumber \\
& + & 2 g^{ab} D_a \phi D_b \phi +  8 (  \lambda - 2 ) ( \lambda -
\frac{3}{2} )  \phi^2 ] \label{220} . \end{eqnarray}

We note that the auxiliary scalar field $\phi$ can be decoupled from
the pure Yang-Mills theory only in the abelian $U(1)$ case and for
the special case $ \lambda = 2 $:

\begin{eqnarray} S(2) & = & - \frac{1}{4 g^2_{YM} } \int d \Omega ( F_{ab}
F^{ab} - 2 \partial_a \phi \partial^a \phi ) \nonumber \\
d \Omega & = & \sqrt{g} d \theta d \phi = sin \theta d \theta d \phi
\ \ \ \ \ \ F^{ab} = g^{a a'} g^{b b'} F_{a' b'} \label{221} .
\end{eqnarray}

When searching a connection between projectors and the matrix model
variable $X_i$, we don't have many choices since the background
matrix $ L_i $  is the only possible definition of derivative, and
we must play  with the vector $ | \psi > $. The natural guess is

\beq X_i = < \psi | L_i | \psi > = L_i + < \psi | [ L_i , | \psi > ]
\label{222} \eeq

which implies the following representation for the fluctuation field
$A_i$ :

\beq A_i = < \psi | [ L_i , | \psi > ] \label{223} \eeq

Since $ [ L_i , . ] \rightarrow_{\alpha \rightarrow 0} - i k_i^a
\partial_a $, we obtain the well-known classical formula for the
monopole connection:

\beq A_i \rightarrow - i k_i^a < \psi | \partial_a | \psi >
\label{224} \eeq

There is however a problem that may ruin the classical limit. If we
use the vectors $ | \psi_{\pm 1 } > $, depending on the oscillator
algebra, rather than the fuzzy sphere algebra, the action of $ L_i $
on $ | \psi_{\pm 1 } > $ is discontinuous in the classical limit.
The only way out is redefining the vectors $ | \psi_{\pm 1 } > $
with an operator acting on the right such that the new vector can be
restricted to a fixed total oscillator number $ \hat{N} = N $.

It turns out that the correction is only possible for $ | \psi_{-1}
> $ using a quasi-unitary operator :

\beq | \psi_{-1} > \rightarrow | \psi_{-1}' > = | \psi_{-1} > U \ \
\ \ U U^{\dagger} = 1 \label{225} \eeq

The quasi-unitary condition keeps the projector $P_{-1}$ invariant.
A possible choice turns out to be :

\beq U_1 = \sum_{n_1, n_2 = 0}^{\infty} | n_1 , n_2 >< n_1+1, n_2 |
\label{226} \eeq

It is not difficult to show that the combination $X_i$ is
proportional to a reducible representation of the Lie algebra :

\begin{eqnarray}
X_i & = & < \psi'_{-1} | L_i | \psi'_{-1} > = \frac{ N+2 }{ N+1 }
U^{\dagger} L_i U \nonumber \\
F_{ij} & = & \frac{ N+2 }{ {( N+1 )}^2 } i \epsilon_{ijk}
U^{\dagger} L_k U  \label{227} \end{eqnarray}

that is indeed a solution of the non-commutative equations of motion
for :

\beq \lambda = 2 + \frac{1}{ N+1 } \label{228} \eeq

This is a direct deformation of the classical monopole solution.
However one can also choose to redefine $ X_i $ as

\beq X_i = U^{\dagger} L_i U \ \ \ \ F_{ij} = 0 \ \ \ \ \lambda = 2
\label{229} \eeq

remembering that in this case the classical limit contains not only
the monopole field but also a constant scalar field, that due to the
$U(1)$ property, remains totally decoupled. In the following we will
shift from one to the other formulation indifferently.

\section{ $U(2)$ projectors and 't Hooft-Polyakov monopoles }

In the case of $ U(1) $ projectors, our strategy was to extend the
well-known classical case, studied in detail in ref. \cite{21}.

When generalizing to the non-abelian 't Hooft-Polyakov monopoles,
the presence of a non-trivial Higgs field complicates the classical
analysis, and it is more convenient starting directly from the
matrix model formalism, which simplifies the whole picture.

We therefore prefer to postulate some non-commutative $ U(2) $
projectors, whose form is dictated by internal consistency, and then
we connect them to known solutions ( see ref. \cite{25} ), leading
to non-commutative extensions of 't Hooft-Polyakov monopoles. The
form of $ U(2) $ projectors, given in terms of oscillators, can be
guessed from the 4d case, where our guide was the classical case of
$ SU(2) $ BPST instantons,  discussed in ref. \cite{24}. This
analogy suggests us to take the following form for $ U(2) $
projectors ( with the simplest topological charge $ k = 1 $ ):

\begin{eqnarray}
P & = & | \psi > < \psi | \nonumber \\
| \psi > &  = & \left( \begin{array}{c} \left(  \begin{array}{cc}
a_0 & -
a^{\dagger}_1 \\ a_1 & a^{\dagger}_0 \end{array} \right) \\
\left(  \begin{array}{cc} a_1 & - a^{\dagger}_0 \\ a_0 &
a^{\dagger}_1 \end{array} \right)
\end{array} \right) f( \hat{N} ) \label{31}
 \end{eqnarray}

It is easy to recognize that to obtain a projector different from
identity, we need to play with the interference between the first
and the second element of the vector $ | \psi > $, since a single
component would produce no functional dependence on the elements of
the fuzzy sphere. Note the exchange of indices ( $ 0 \leftrightarrow
1 $ ) in the second element of the vectors which is responsible for
a non-trivial interference.

We will notice in the following that adding the second element with
interchanged indices is also important to find a simple result for
the expectation value $ < \psi | L_i | \psi > $, the form which has
enabled us to derive the non-commutative abelian monopoles in the
matrix model formalism.

As in the $U(1)$ case, the function $ f( \hat{N} )$ can be
determined by imposing the normalization condition on $ | \psi > $:

\beq < \psi | \psi > = f^2 ( \hat{N} ) \left(
\begin{array}{cc} 2 \hat{N} & 0 \\ 0 & 2 ( \hat{N} + 2 ) \end{array}
\right) \label{32} \eeq

Therefore the form of $ f ( \hat{N} ) $ is fixed as the following
diagonal form :

\beq f ( \hat{N} ) = \left(
\begin{array}{cc} \frac{1}{\sqrt{2 \hat{N}}} & 0 \\ 0 & \frac{1}{\sqrt{2 ( \hat{N} + 2
)}} \end{array} \right) \label{33} \eeq

Using the commutation rule of the oscillators, the final form of the
vector $ | \psi > $ can be simplified as :

\beq | \psi >   = \frac{1}{\sqrt{ 2 ( \hat{N} + 1 )}} \left(
\begin{array}{c} \left(
\begin{array}{cc} a_0 & -
a^{\dagger}_1 \\ a_1 & a^{\dagger}_0 \end{array} \right) \\
\left(  \begin{array}{cc} a_1 & - a^{\dagger}_0 \\ a_0 &
a^{\dagger}_1 \end{array} \right)
\end{array} \right)  \label{34}
\eeq

The corresponding non-commutative $ U(2) $ projector, that will be
further elaborated for a possible connection with the 't
Hooft-Polyakov monopoles, is given by :

\beq P = | \psi > < \psi | = \frac{1}{ 2 ( \hat{N} + 1 )} \left(
\begin{array}{cc}
\left(  \begin{array}{cc} \hat{N} + 1  & 0 \\ 0 & \hat{N} + 1
\end{array} \right) &
\left( \begin{array}{cc} 2 a_0 a^{\dagger}_1 &
  a_0 a^{\dagger}_0 - a^{\dagger}_1 a_1  \\
  a_1 a^{\dagger}_1 - a^{\dagger}_0 a_0  &
 2 a^{\dagger}_0 a_1 \end{array}  \right)       \\
\left(   \begin{array}{cc} 2 a^{\dagger}_0 a_1 &   a_1 a^{\dagger}_1
- a^{\dagger}_0 a_0
\\  a_0 a^{\dagger}_0 - a^{\dagger}_1 a_1
 & 2 a_0 a^{\dagger}_1
\end{array} \right) & \left( \begin{array}{cc}  \hat{N} + 1  & 0 \\ 0 &
\hat{N} + 1
\end{array} \right)
\end{array}
\right) \label{35} \eeq

At this level, it is safe substituting to the number operator $
\hat{N} $ its eigenvalue $ N $, and considering the entries of $ P $
as elements of the fuzzy sphere function space.

The trace of the projector is obviously :

\beq Tr P = 2 \ Tr I < Tr 1_P = 4 Tr I \label{36} \eeq

Being this projector non-trivial, it can be taken as a natural
candidate for a non-abelian topological configuration. To be sure,
we must study its connection with matrix models.

We already know what can be expected from our previous study of $
U(1) $ projectors. The combined presence of operators of type $ a (
a^{\dagger} ) $ implies that this projector will act on the
background in order to increase ( decrease ) the dimension of the
representation used in the background. This is exactly the
characteristic of a topologically non-trivial configuration that in
ref. \cite{25} has been shown to reproduce in the classical limit
the 't Hooft-Polyakov monopoles. We feel therefore to be on the
right way and the strategy of projectors combined with the use of
matrix models can also, as a byproduct, teach us how to treat the
non-abelian topology on the classical sphere, generalizing the
mathematical work of \cite{21}.

\section{Connection with matrix models}

Tentatively, we can try to connect the projectors with matrix
models, as successfully done in the $ U(1) $ case,

\beq X_i = < \psi | L_i | \psi > \label{41} \eeq

However we easily recognize that, in the classical limit, this
formula is inconsistent with the presence of a non-trivial Higgs
field, since the $ L_i $ action reduces to $ k_i^a \partial_a $ and
therefore it projects in the tangent plane to the sphere, while the
Higgs field fluctuation is in the orthogonal direction, along the
radius. The explicit calculation of the matrix element $ < \psi |
L_i | \psi > $ will suggest us what we need to add to this formula
to complete the connection with matrix models in the non-abelian
case.

The explicit calculation gives, in details,

\beq < \psi | L_i | \psi > = \frac{1}{2} \left(
\begin{array}{cc} \frac{1}{\hat{N}} ( a^{\dagger}_0 L_i a_0 +
a^{\dagger}_1 L_i a_1  ) & \frac{1}{\hat{N}} ( - a^{\dagger}_0 L_i
a^{\dagger}_1 + a^{\dagger}_1 L_i a^{\dagger}_0 ) \\
\frac{1}{\hat{N}+ 2} ( - a_1 L_i a_0 + a_0 L_i a_1 ) &
\frac{1}{\hat{N}+ 2} ( a_0 L_i a^{\dagger}_0 + a_1 L_i a^{\dagger}_1
)
\end{array} \right) + ( 0 \leftrightarrow 1 ) \label{42}
 \eeq

We note from this formula the importance of the contribution ( $ 0
\leftrightarrow 1 $ ) to cancel the off-diagonal terms.

In summary we obtain:

\beq < \psi | L_i | \psi > =  \left(
\begin{array}{cc} \frac{1}{\hat{N}} ( a^{\dagger}_0 L_i a_0 +
a^{\dagger}_1 L_i a_1  ) & 0  \\
0  & \frac{1}{\hat{N}+ 2} ( a_0 L_i a^{\dagger}_0 + a_1 L_i
a^{\dagger}_1 )
\end{array} \right) \label{43} \eeq

It is not difficult to compute the terms inside the parenthesis
using the oscillator algebra:

\beq < \psi | L_i | \psi > =  \left(
\begin{array}{cc} \frac{\hat{N}-1}{\hat{N}} L_i & 0  \\
0  & \frac{\hat{N} + 3}{\hat{N} + 2}  L_i \end{array} \right)
\label{44} \eeq

As we discussed in our previous article \cite{23}, the action of $
L_i $ on $ \psi $ cannot be smoothly connected to the classical Lie
derivative on the sphere unless we project the vector $ | \psi >$ on
the fuzzy sphere algebra. The only possibility left, with the
constraint of keeping invariant the projector $ P $, is dressing $ |
\psi > $ with a quasi-unitary operator, such that $ | \psi' > $
belongs to the fuzzy sphere function space:

\begin{eqnarray}
P & = & | \psi > < \psi | = | \psi' > < \psi' | \nonumber \\
| \psi' > & = & | \psi > U \label{45} \end{eqnarray}

The quasi-unitary operator $ U $, as in the case of non-commutative
extension of Dirac monopoles, plays an essential role to define the
non-abelian topology. The only consistent choice, apart from unitary
gauge transformation, given the structure of the vector $ | \psi >
$, is the following quasi-unitary operator :

\begin{eqnarray}
U & = & \left( \begin{array}{cc} U_1^{\dagger} & U_{12}^{\dagger} \\
0 & U_2 \end{array}\right) \nonumber \\
U_1 & = & \sum_{n_1, n_2 =0}^{\infty} | n_1, n_2 > < n_1 + 1, n_2 |
\nonumber \\
U_2 & = & \sum_{n_1, n_2 =0}^{\infty} | n_1, n_2 > < n_1, n_2 + 1|
\nonumber \\
U_{12} & = & \sum_{n_1=0}^{\infty} | n_1, 0 > < 0 , n_1 + 1|
\label{46} \end{eqnarray}

This operator satisfies to the following properties:

\beq U U^{\dagger} = \left( \begin{array}{cc} U_1^{\dagger} & U_{12}^{\dagger} \\
0 & U_2 \end{array} \right)
\left( \begin{array}{cc} U_1 & 0 \\
U_{12} & U^{\dagger}_2 \end{array} \right) = \left(
\begin{array}{cc} U_1^{\dagger} U_1 + U_{12}^{\dagger} U_{12} = 1 -
|0><0| & U^{\dagger}_{12} U^{\dagger}_2 = 0 \\ U_2  U_{12} = 0 & U_2
U_2^{\dagger} = 1
\end{array} \right) \label{47} \eeq

Since the operator $ | 0 > < 0 | $ is annihilated by the action of $
a_0 $ and $ a_1 $, the combination $ U U^{\dagger} $ behaves as the
identity when acting on the oscillators.

Moreover the following property holds :

\beq U^{\dagger} U =  \left( \begin{array}{cc} U_1 & 0 \\
U_{12} & U^{\dagger}_2 \end{array} \right)
\left( \begin{array}{cc} U_1^{\dagger} & U_{12}^{\dagger} \\
0 & U_2 \end{array} \right) = \left(
\begin{array}{cc} U_1 U_1^{\dagger} = 1
 & U_1 U_{12}^{\dagger} = 0 \\ U_{12} U_1^{\dagger} = 0 &
U_2^{\dagger} U_2 + U_{12} U_{12}^{\dagger} = 1 \end{array} \right)
 \label{48} \eeq

This operator doesn't change the rank of the background but simply
it changes the dimensions of the component representations.
Therefore we must expect that :

\beq \left( \begin{array}{cc} ( L_i )_{N+1} & 0 \\
0 & ( L_i )_{N+1} \end{array} \right) \rightarrow \left[ U^{\dagger}
\left( \begin{array}{cc} L_i  & 0 \\
0 & L_i  \end{array} \right) U \right]_{N+1} = \left(
\begin{array}{c|c} {( L_i )}_{N+2} & 0 \\ \hline 0 & {( L_i )}_{N}
\end{array} \right) \label{49} \eeq

where the basic building blocks in the last formula have different
size from those of the background.

Now let's apply $ | \psi ' > $ to the formula(\ref{41}). In this
case we obtain

\begin{eqnarray}
X_i & = & < \psi' | L_i | \psi' > = U^{\dagger}
\left( \begin{array}{cc} \frac{\hat{N}-1}{\hat{N}} L_i & 0 \\
0 & \frac{\hat{N} + 3}{\hat{N} + 2} L_i  \end{array} \right) U =
\nonumber \\
& = & \frac{\hat{N}}{\hat{N}+1} \left( \begin{array}{cc} U_1 L_i
U_1^{\dagger} & U_1 L_i U_{12}^{\dagger} \\
U_{12} L_i U_1^{\dagger} & U_{12} L_i U_{12}^{\dagger}
\end{array} \right) +
\frac{\hat{N}+2}{\hat{N}+1} \left( \begin{array}{cc} 0  & 0 \\
0 & U^{\dagger}_{2} L_i U_2 \end{array} \right) \label{410}
\end{eqnarray}

It is not difficult to recognize that this formula is nothing but a
sum of $ SU(2) $ representations ( see Appendix ):

\beq \left( \begin{array}{cc} U_1 L_i
U_1^{\dagger} & U_1 L_i U_{12}^{\dagger} \\
U_{12} L_i U_1^{\dagger} & U_{12} L_i U_{12}^{\dagger}
\end{array} \right)_{N+1} = \left(
\begin{array}{c|c} {( L_i )}_{N+2} & 0 \\ \hline 0 & 0
\end{array} \right) \label{411}
\eeq

while

\beq \left( \begin{array}{cc} 0  & 0 \\
0 & U^{\dagger}_{2} L_i U_2 \end{array} \right)_{N+1} = \left(
\begin{array}{c|c} 0 & 0 \\ \hline 0 & {( L_i )}_N
\end{array} \right) \label{412} \eeq

In summary, the presence of the quasi-unitary operator $ U $ can be
completely worked out into this final formula :

\beq X_i = < \psi' | L_i | \psi' > =  \left(
\begin{array}{c|c} \frac{\hat{N}}{\hat{N}+1} {( L_i )}_{N+2} & 0 \\
\hline 0 & \frac{\hat{N}+2}{\hat{N}+1} {( L_i )}_{N}
\end{array} \right) \label{413} \eeq

This block-diagonal form is still not an explicit solution of the
equations of motion. The nearest solution is very simple to obtain,
redefining $ | \psi ' > $ with a diagonal matrix :

\beq X_i = \left( \begin{array}{cc} f_{+} + f_{-} & 0 \\
0 & f_{+} - f_{-} \end{array} \right) < \psi' | L_i | \psi' >
\left( \begin{array}{cc} f_{+} + f_{-} & 0 \\
0 & f_{+} - f_{-} \end{array} \right) \label{414} \eeq

This final form is not of type $ < \psi' | L_i | \psi' > $, as in
the case of Dirac monopoles; however as we discussed in the
beginning, this modification is necessary to obtain in the classical
limit a non-trivial contribution for the Higgs field.

Finally requiring that

\beq X_i =   \left(
\begin{array}{c|c}  {( L_i )}_{N+2} & 0 \\
\hline 0 &  {( L_i )}_{N}
\end{array} \right) \label{415}\nonumber \eeq

we obtain a condition that fixes the unknown constants $ f_{+},
f_{-} $:

\beq f_{\pm} = \frac{1}{2} \left[ \sqrt{ \frac{ \hat{N} + 1}{
\hat{N} }} \pm \sqrt{ \frac{ \hat{N} + 1}{ \hat{N} + 2 }} \right]
 \label{416} \eeq

This solution can also be expressed, with a gauge transformation,
as:

\beq X_i = L_i \otimes 1 + 1 \otimes S_i \label{417}  \eeq

Now it is simpler to extract the contribution of the fluctuation (
non-abelian monopole field ) from the background:

\beq X_i - ({ \rm background }) = 1 \otimes S_i \label{418} \eeq

An alternative way to express formula (\ref{414}), reaching the same
solution, is redefining both $ | \psi' > $ and the projector $ P $
as

\begin{eqnarray}
| \psi'' > & = & \left( \begin{array}{cc} \frac{1}{f_{+} + f_{-}} & 0 \\
0 & \frac{1}{f_{+} - f_{-}} \end{array} \right) | \psi' >
\left( \begin{array}{cc} f_{+} + f_{-} & 0 \\
0 & f_{+} - f_{-} \end{array} \right) \nonumber \\
P' & = & | \psi'' > < \psi'' | = \left( \begin{array}{cc} \frac{1}{f_{+} + f_{-}} & 0 \\
0 & \frac{1}{f_{+} - f_{-}} \end{array} \right) P
\left( \begin{array}{cc} f_{+} + f_{-} & 0 \\
0 & f_{+} - f_{-} \end{array} \right)
 \label{419} \end{eqnarray}

The formula, analogous to (\ref{414}), linking projectors to
connections, can be expressed in this case as

\begin{eqnarray}  X_i & = &
< \psi'' | \left( \begin{array}{cc} f_{+} + f_{-} & 0 \\
0 & f_{+} - f_{-} \end{array} \right) L_i
\left( \begin{array}{cc} f_{+} + f_{-} & 0 \\
0 & f_{+} - f_{-} \end{array} \right) | \psi'' >
 \nonumber \\
& = & < \psi'' | \left( \begin{array}{cc} \frac{ \hat{N} +
1}{\hat{N}} L_i & 0 \\ 0 & \frac{ \hat{N} + 1}{\hat{N}+ 2} L_i
\end{array} \right) | \psi'' > = < \psi'' | X^{0}_i | \psi''
> \label{420} \end{eqnarray}

This version allows us to define a gauge invariant version of matrix
models, generalizing what we have done in the $ U(1) $ case ( ref.
\cite{23} ), built directly on the projectors

\beq X_i = P' X^{0}_i P' \label{421}\eeq

that, by construction, satisfies the same equations of motion.

\section{ Classical limit }

The lagrangian of the matrix model ( eq. (\ref{214}) ) for $ \lambda
= 2 $ in the non-abelian case leads to the following classical
action on the sphere:

\beq S = - \frac{1}{ 4 g^2_{YM} }\int d \Omega ( F_{ab} F_{a'b'}
g^{aa'} g^{bb'} + 2 g^{aa'} D_a \phi D_{a'} \phi ) \label{51} \eeq

where we define

\begin{eqnarray}
D_a \phi & = & - i \partial_a \phi + [ A_a, \phi ] \nonumber \\
F_{ab} & = & - i ( \partial_a A_b - \partial_b A_a ) + [ A_a , A_b ]
 \label{52} \end{eqnarray}

The variation with respect to $ A_a $ and $ \phi $ is vanishing by
simply requiring that

\beq D_a ( \sqrt{ g } g^{aa'} g^{bb'} F_{a' b'} ) = D_a \phi = 0
\label{53}\eeq

On the other hand, we know that at non-commutative level

\beq X_i = L_i \otimes 1 + 1 \otimes S_i \label{54}\eeq

satisfies the constraint

\beq F_{ij} = [ X_i, X_j ] - i \epsilon_{ijk} X_k = 0 \label{55}
\eeq

since $ X_i $ satisfies the commutation rules of the angular
momentum. In the classical limit the fluctuation field $ A_i $

\beq A_i = 1  \otimes S_i = k_i^a A_a + n_i \phi \label{56} \eeq

can be projected over the tangent plane of the sphere and on the
orthogonal direction

\begin{eqnarray}
\phi & = & n_i \otimes S_i \nonumber \\
A_a & = & g_{ab} \ k_i^b \otimes S_i
 \label{57} \end{eqnarray}

We must obtain, as a check, that

\begin{eqnarray}
F_{ij} & = & \frac{1}{R} \epsilon_{ijk} x_k \left( i \phi +
\frac{\epsilon^{ab}}{ 2 \sqrt{g}} F_{ab} \right) + \frac{1}{R} x_j
k_i^a D_a \phi - \frac{1}{R} x_i k_j^a D_a \phi = 0 \nonumber \\
D_a \phi & = & 0 \nonumber \\
F_{ab} & = & - i \epsilon_{ab} \sqrt{g} \phi \label{58}
\end{eqnarray}

We note that these constraints are enough to solve the classical
equations of motion ( eq. (\ref{53}) ) for $ \lambda = 2 $. This
check permits us to have an explicit formula for $ \phi $ and $ A_a
$:

\begin{eqnarray}
\phi & = & n_i \otimes S_i = \frac{1}{2} \left(
\begin{array}{cc} cos \theta & sin \theta e^{- i \phi } \\
sin \theta e^{i \phi} & - cos \theta \end{array}
\right)\nonumber \\
A_\theta & = & k_i^\theta \otimes S_i = \frac{i}{2} \left(
\begin{array}{cc} 0 & - e^{- i \phi } \\
e^{i \phi} & 0 \end{array}
\right)\nonumber \\
A_\phi & = & sin^2 \theta \ k_i^\phi \otimes S_i = \frac{sin
\theta}{2} \left(
\begin{array}{cc} sin \theta & - cos \theta e^{- i \phi } \\
- cos \theta e^{i \phi} & - sin \theta \end{array} \right)
 \label{59}\end{eqnarray}

It is easy to deduce that

\beq F_{\theta\phi} = - \frac{i}{2} ( \partial_\theta A_\phi -
\partial_\phi A_\theta ) = - [ A_\theta , A_\phi ] = - i \sqrt{g}
\phi \label{510} \eeq

that is equivalent to eq. (\ref{58}), with the notation $
\epsilon_{\theta\phi} = 1 $.

Moreover

\begin{eqnarray}
- i \partial_\theta \phi & = & - [ A_\theta, \phi ] = - \frac{i}{2}
\left(
\begin{array}{cc} - sin \theta & cos \theta e^{- i \phi } \\
cos \theta e^{i \phi} & sin \theta \end{array} \right) \nonumber \\
- i \partial_\phi \phi & = & - [ A_\phi, \phi ] =
\frac{sin\theta}{2} \left(
\begin{array}{cc} 0 & -  e^{- i \phi } \\
 e^{i \phi} & 0 \end{array} \right) \label{511}
\end{eqnarray}

The classical limit induced by our formula (\ref{414}) produces a
slightly different solution. Starting from the classical limit of
the vector $ | \psi' > $:

\beq | \psi' > \rightarrow_{N \rightarrow \infty} | \psi_{cl} >  =
\frac{1}{\sqrt{2}} \left(
\begin{array}{c} \left(
\begin{array}{cc} a_0 & -
a^{*}_1 \\ a_1 & a^{*}_0 \end{array} \right) \\
\left(  \begin{array}{cc} a_1 & - a^{*}_0 \\ a_0 & a^{*}_1
\end{array} \right)
\end{array} \right) \ \ \ \ \ < \psi_{cl} | \psi_{cl} > = 1
\label{512}\eeq

we obtain that the variable $ X_i $ of the classical matrix model is
of the type:

\beq X_i \simeq ( f_{+}^2 + f_{-}^2 ) < \psi_{cl} | L_i | \psi_{cl}
> + 2 f_{+} f_{-} \left(
\begin{array}{cc} 1 & 0 \\ 0 & -1 \end{array} \right) < \psi_{cl} | L_i
| \psi_{cl} > \label{513} \eeq

Since

\beq f_{+} \rightarrow 1 \ \ \ \ f_{-} \rightarrow \frac{1}{2 N}
\label{514}\eeq

we can deduce that

\beq X_i \simeq < \psi_{cl} | L_i | \psi_{cl} > +  \frac{1}{2}
\left( \begin{array}{cc} 1 & 0 \\ 0 & -1 \end{array} \right) <
\psi_{cl} | \frac{\hat{x}_i}{R} | \psi_{cl} > \label{515}\eeq

where $ \hat{x}_i = \alpha L_i $ is the operator deforming the
coordinate $x_i$. The Higgs field contribution comes exactly from
this last term:

\begin{eqnarray}
\phi & = & \frac{1}{2}  \left(
\begin{array}{cc} 1 & 0 \\ 0 & -1 \end{array} \right) \nonumber \\
A_{\theta} & = & - i < \psi_{cl} | \partial_{\theta} | \psi_{cl} > =
0
\nonumber \\
A_{\phi} & = & - i < \psi_{cl} | \partial_{\phi} | \psi_{cl} > = -
\frac{ cos\theta}{2} \left(
\begin{array}{cc} 1 & 0 \\ 0 & -1 \end{array} \right)
\label{516} \end{eqnarray}

with the choice

\begin{eqnarray} a_0 & = & cos \frac{\theta}{2}
e^{-i \frac{\phi}{2}} \nonumber \\
a_1 & = & sin \frac{\theta}{2} e^{i \frac{\phi}{2}} \label{518}
\end{eqnarray}

Naturally there exists a classical gauge transformation that
connects the solution (\ref{59}) to the solution (\ref{516})

\begin{eqnarray}
\phi_g & = & n_i \otimes  S_i \ \ \ \ \ \  \ \ \ \ \ \ \ \phi  = S_3 \nonumber \\
A_{ag} & = & g_{ab} k_i^b \otimes S_i \ \ \ \ \ \ \ \ \ \ A_\theta
= 0 \ \ A_\phi = - cos \theta S_3 \label{519}\end{eqnarray}

In both cases the constraint $ F_{ij} = 0 $ is satisfied. Starting
for example from the Higgs field :

\beq \phi_g = g^{-1} \phi g = \frac{1}{2} \left( \begin{array}{cc}
\overline{a} & - b \\ \overline{b} & a \end{array} \right) \left(
\begin{array}{cc} 1 & 0 \\ 0 & - 1 \end{array}
\right) \left( \begin{array}{cc} a & b \\ - \overline{b} &
\overline{a}
\end{array} \right)
\ \ \ \ \ \ a \overline{a} + b \overline{b} = 1
 \label{520}\eeq

the gauge transformation is defined by :

\begin{eqnarray}
a \overline{a} - b \overline{b} &  =& cos \theta \nonumber \\
2 \overline{a} b & = & sin \theta e^{ - i \phi } \nonumber \\
a \overline{a} + b \overline{b} & = & 1 \label{521} \end{eqnarray}

which is solved by

\begin{eqnarray}
a & = & cos \frac{\theta}{2} e^{ i \frac{\phi}{2}} \nonumber \\
b & = &  sin \frac{\theta}{2} e^{ - i \frac{\phi}{2}} \label{522}
\end{eqnarray}

We have checked that $ A_{ag} $ and $ A_a $ are related by the same
gauge transformation.

\section{ Deforming the Chern Class}

In literature a candidate for an eventual non-commutative
topological index has been proposed, by using an action taking
values only in the integer numbers \cite{25}. In this paper we want
to suggest an alternative, more traditional, definition, by taking
an action taking values not necessarily in the integers.

Our candidate for the non-commutative Chern class is the
Chern-Simons term

\beq S_{CS} = \frac{1}{N+1} Tr \left[ \frac{2i}{3} \epsilon_{ijk}
X_i X_j X_k + X_i X_i \right] \label{61} \eeq

This action, as we have found in a previous paper ( ref. \cite{26}
), is invariant under deformed diffeomorphisms, a property which
doesn't hold for the standard Yang-Mills action.

We want to show that this action in the classical limit, once
evaluated on the 't Hooft-Polyakov monopole, corresponds to the
classical topological number

\beq Q = - \frac{1}{8\pi} \int_S d \Omega \epsilon_{ijk} n_i
\epsilon^{abc} n^a (
\partial_j n^b ) ( \partial_k n^c ) = - \frac{1}{4\pi} \int_S d
\Omega = - 1 \label{62}\eeq

Comparing with ref. \cite{25}, the integral $ Q $ is equivalent to
the following action

\beq Q = \frac{1}{8\pi} Tr \int_S d \Omega \epsilon_{ijk} n_i
F_{jk}^{ \perp } \phi \label{63} \eeq

where

\begin{eqnarray}
\phi & = & n_i \otimes S_i \nonumber \\
 F_{ij}^{ \perp } &  =  & \partial_i
a_j^{ \perp } -
\partial_j a_i^{ \perp } - i [ a_i^{ \perp } , a_j^{ \perp } ]
\label{64}
 \end{eqnarray}

and $ a_i^{\perp} $ is the orthogonal part of the fluctuation $ A_i
$, i.e. in our notations

\beq a_i^{\perp} = \epsilon_{ijk} n_j  A_k \label{65}\eeq

Extracting the contribution of the transverse part and substituting
the explicit relation between $ F_{ab} $ and $ \phi $ given by
(\ref{510})

\beq \epsilon_{ijk} n_i F_{ij}^{\perp} = - 2 \phi = -i  \left(
\frac{\epsilon^{ab}}{\sqrt{g}} F_{ab} \right)\label{66}\eeq

we can prove that

\beq Q = \frac{1}{8\pi}Tr \int_S d \Omega \epsilon_{ijk} n_i \
F_{jk}^{\perp} \phi  = -  \frac{1}{8\pi} Tr \int_S d \Omega \left[ \
i \phi \frac{\epsilon^{ab}}{\sqrt{g}} F_{ab} \ \right]\label{67}
\eeq

This expression is very similar to the classical limit of the
Chern-Simons topological term (\ref{61}). In fact, from references
\cite{10} and \cite{12} we obtain that

\beq S_{CS} \rightarrow_{\alpha \rightarrow 0} \frac{1}{8\pi} Tr
\int d \Omega \left( i \phi \frac{\epsilon^{ab}}{\sqrt{g}} F_{ab} -
\phi^2 \right) \label{68} \eeq

Applying again the classical limit of the constraint $ F_{ij} = 0 $
i.e.

\begin{eqnarray}
F_{ab} & = & - i \epsilon_{ab} \sqrt{g} \phi \nonumber \\
S & \rightarrow_{\alpha \rightarrow 0}&  \frac{1}{8\pi} Tr \int d
\Omega \left( i \phi \frac{\epsilon^{ab}}{2 \sqrt{g}} F_{ab} \right)
= - \frac{Q}{2} = \frac{1}{2} \label{69}
\end{eqnarray}

and substituting the explicit value for $ \phi $ given by eq.
(\ref{59}) we obtain perfect agreement.

We have proved that, at least in the classical limit, the
Chern-Simons action, evaluated on the classical solution (\ref{56}),
produces the topological number characteristic of non-abelian
monopoles.

Now we want to show that the same topological number can be obtained
evaluating the Chern-Simons action at pure non-commutative level:

\beq S_{CS}^{\rm monopole} - S_{CS}^{\rm bg} = -\frac{Q}{2} =
\frac{1}{2} \label{610}\eeq

and that in the non-abelian case there is a continuity with the
non-commutative extension. If we evaluate $ S_{CS} $ on the
background we obtain :

\beq S_{CS}^{\rm bg} = \frac{1}{3 ( N+1 )} Tr \left(
\begin{array}{cc} ( L_i L_i )_{N+1} & 0 \\ 0 & ( L_i L_i )_{N+1}
\end{array} \right)= \frac{N(N+2)}{6} \label{611} \eeq

while on the monopole configuration

\beq S_{CS}^{\rm monopole} = \frac{1}{3 ( N+1 )} Tr \left(
\begin{array}{c|c} ( L_i L_i )_{N+2} & 0 \\ \hline 0 & ( L_i L_i )_{N}
\end{array} \right) = \frac{ N^2 + 2 N + 3}{6} \label{612} \eeq

Therefore

\beq S_{CS}^{\rm monopole} - S_{CS}^{\rm bg} = \frac{1}{8\pi} \int_S
Tr \phi^2 = \frac{1}{2} \label{613} \eeq

the classical topological number is maintained by the
non-commutative deformation in the non-abelian case.

To complete the picture, we want to analyze what happens for  Dirac
monopoles. At first sight, the classical limit of the Chern-Simons
term doesn't seem to be in agreement with the Chern class

\beq S_{Ch} = \int_S d \Omega \left( i \frac{ \epsilon^{ab} }{
\sqrt{g} } F_{ab} \right) = Q \label{614} \eeq

because of the absence of the Higgs field. This problem can be
easily circumvented by admitting that the classical monopole
configurations have also a constant Higgs field. Since the
lagrangian of the Higgs field is decoupled, this configuration is
still solution of the equations of motion.

Therefore the evaluation of $S_{CS}$ is equivalent to the evaluation
of $S_{Ch}$, by allowing the presence of a constant Higgs field.
This different classical limit can be realized starting from

\beq X_i^{NC} = U^{\dagger} L_i U \label{615} \eeq

instead of

\beq X_i^{NC} = \frac{ N+2 }{ N+1 } \ U^{\dagger} L_i U \label{616}
\eeq

This new solution satifies the same commutation rules of the angular
momentum and therefore to the constraint

\beq F_{\theta\phi} = - i \partial_{\theta} A_{\phi} = i
\epsilon_{\theta\phi} \sqrt{g} \phi \leftarrow_{\alpha \rightarrow
0} \ F_{ij} = 0 \label{617} \eeq

and the explicit solution is

\beq X_i^{NC} \rightarrow_{\alpha \rightarrow 0} \left\{
\begin{array}{c} A_\phi = \frac{ cos \theta - 1 }{2} \\ \phi = - \frac{1}{2}
\end{array} \right. \label{618} \eeq

In summary the classical fluctuation that must be compared with the
solution $X_i^{NC}$ is of the type:

\begin{eqnarray}
a_i & = & k_i^a A_a + \frac{x^i}{R} \phi \nonumber \\
k_i^a A_a & = & k_i^\phi A_\phi = - \frac{1}{2} \left[
\frac{cos\theta}{sin\theta} ( cos \theta - 1 ) cos \phi ,
\frac{cos\theta}{sin\theta} ( cos \theta - 1 ) sin \phi, 1 - cos
\theta \right] \nonumber
\\
n^i \phi & = & - \frac{1}{2} [ sin \theta cos \phi, sin \theta sin
\phi, cos
\theta ] \nonumber \\
a_i & \sim & - \frac{1}{2} \left[ \frac{1 - cos \theta}{ sin \theta
} cos \phi, \frac{1 - cos \theta}{ sin \theta } sin \phi, 1 \right]
\simeq ( L_i )_N - ( L_i )_{N+1} \label{619} \end{eqnarray}

The direct non-commutative evaluation in this case produces a
singularity:

\begin{eqnarray}
S_{CS}^{\rm bg} & = & \frac{1}{3 ( N+1 )} Tr ( L_i L_i )_{N+1} =
\frac{N ( N+2 )}{12} \nonumber \\
S_{CS}^{\rm monopole} & = & \frac{1}{3 ( N+1 )} Tr \left(
\begin{array}{cc}
( L_i L_i )_{N} & 0 \\ 0 & 0 \end{array} \right) = \frac{N(N-1)}{12}
\nonumber \\
S_{CS}^{\rm monopole} &  - & S_{CS}^{\rm bg} = \frac{N(N-1)}{12} -
\frac{N ( N+2 )}{12} = - \frac{N}{4} \label{620} \end{eqnarray}

This negative result requires to analyze more carefully the
classical limit of the Dirac monopoles. Starting from the components
$ a_{\pm}$

\begin{eqnarray} a_{+} & = & ( U^{\dagger} L_{+} U )_{N+1} - ( L_{+} )_{N+1} =
\sum ( \sqrt{ n_1 } - \sqrt{ n_1 + 1 } ) \sqrt{ n_2 + 1 } | n_1 + 1,
n_2 > < n_1 , n_2 + 1 | \nonumber \\
& \simeq & - \frac{1}{2} \sum \sqrt{ \frac{ n_2+1 }{ n_1+1 } } |
n_1+1, n_2 >< n_1, n_2+1 | = - \frac{1}{2} \frac{1}{\hat{z}_1}
\hat{z}_2
 \label{621} \end{eqnarray}

where we have defined

\begin{eqnarray} \frac{1}{\hat{z}_1} & = & \sum_{n_1}
\frac{1}{\sqrt{n_1+1}} | n_1+1 ><n_1 | \ \ \ \ \ \ \ \ z_1 =
\sqrt{2} \ cos
\frac{\theta}{2} e^{-i \frac{\phi}{2}} \nonumber \\
\hat{z}_2 & = & \sum_{n_2} \sqrt{ n_2 + 1 } | n_2 >< n_2 + 1 | \ \ \
\ \ \ \ \ \ \ z_2 = \sqrt{2} \ sin \frac{\theta}{2} e^{i
\frac{\phi}{2}} \nonumber \\
x_{+} & = & \overline{z}_1 z_2 = sin \theta e^{ i \phi } \ \ \ \ x_3
= cos \theta \nonumber \\
a_{+} & \rightarrow & - \frac{1}{2 z_1} z_2 = - \frac{1}{2} \frac{
sin \frac{\theta}{2} }{ cos \frac{\theta}{2} } e^{ i \phi } =
\frac{1}{2} \frac{ cos \theta - 1 }{ sin \theta } e^{ i \phi }
\label{622}
\end{eqnarray}

we find that $ a_{+} $ and $ a_{-} $ are really continuous
deformations of the classical configurations. Different is the case
of $ a_3 $:

\begin{eqnarray}
a_3 & = & ( U^{\dagger} L_3 U )_{N+1} - ( L_3 )_{N+1} = \frac{1}{2}
\left( - \sum_{n_1 \neq 0} | n_1, n_2 >< n_1, n_2 | + \sum_{n_2} n_2
| 0, n_2
><0, n_2 |
\right)_{N+1} = \nonumber \\
& = & - \left( \frac{1 - P_0}{2} \right) + \frac{ N+1 }{2} P_0 \ \ \
\ \ P_0 = | 0, N+1
>< 0, N+1 | \label{623} \end{eqnarray}

$ a_3 $ has a problem, i.e. a discontinuous term that has support on
a single state:

\begin{eqnarray}
a_3^m & = & - \left(  \frac{ 1 - P_0 }{2} \right) \nonumber \\
a_3^s & = & \frac{N+1}{2} P_0
 \label{624} \end{eqnarray}

The effect of $ a^s_3 $ seems at first sight negligible, if we
compare a single state with respect to an infinite number. However
on the non-commutative action it makes a difference and produces a
nasty discontinuity with the classical topological number. Let us
try to exclude the contribution of this state, redefining the
solution as

\begin{eqnarray}
X_i & = & ( L_i + a_i^s ) + a_i^m \nonumber \\
X'_i & = & X_i - a_i^s = L_i + a_i^m \nonumber \\
S_{CS}^{\rm monopole} - S_{CS}^{\rm bg} & = & S_{X'} - S_{L_i} =
\frac{ N^2+2N+3 }{12} - \frac{N(N+2)}{12}  = \frac{1}{4} = \nonumber \\
& = & \frac{1}{4\pi} \int_S d \Omega \  \left( i \phi \frac{
\epsilon^{ab}}{2 \sqrt{g}} F_{ab} \right) = - \frac{Q}{4} =
\frac{1}{4}
 \label{625} \end{eqnarray}

The agreement now is perfect.

\section{Conclusions}

In this work we have shown how to characterize easily the
topologically nontrivial configurations leading to 't Hooft-Polyakov
monopoles with a noncommutative $ U(2) $ projector. The
non-triviality of the $U(2)$ projector is assured in the classical
limit by the use of the Hopf fibration $ \pi : S^3 \rightarrow S^2
$, since the projector cannot be decomposed in terms of vectors
belonging to the sphere $S^2$ but only to $S^3$.

At noncommutative level we can surely state that there exist
solutions to the matrix model equations of motion extending in a
smooth way classical topology.

It remains an open question how to characterize the topological
meaning of these noncommutative configurations, whether they remain
stable and so on.

As a first step in this direction we have suggested a candidate for
deforming the Chern class, maintaining invariant the classical
topological number. However, this subject still requires a deeper
investigation in the future.

\appendix
\section{Appendix}

Aim of this appendix is showing that the action of the quasi-unitary
operator produces the following modification of the representations
of the Lie algebra:

\beq ( U^{\dagger} L_i U )_{N+1} = \left[ \left( \begin{array}{cc}
U_1 & 0
\\ U_{12} & U^{\dagger}_2 \end{array} \right) L_i
\left( \begin{array}{cc} U_1^{\dagger} & U_{12}^{\dagger}
\\ 0 & U_2 \end{array} \right) \right]_{N+1} = \left(
\begin{array}{c|c} (L_i)_{N+2} & 0 \\ \hline 0 & (L_i)_N \end{array}
\right) \label{a1}\eeq

where

\begin{eqnarray}
U_1 & = & \sum_{n_1, n_2} | n_1, n_2 >< n_1 +1, n_2 | \nonumber \\
U_2 & = & \sum_{n_1, n_2} | n_1, n_2 >< n_1, n_2 +1 | \nonumber \\
U_{12} & = & \sum_{n_2} | n_1 ,  0 >< 0, n_1 +1 | \label{a2}
\end{eqnarray}

We firstly observe that the following part is trivially verified

\beq \left[ \left( \begin{array}{cc} 0 & 0
\\ 0 & U^{\dagger}_2 \end{array} \right) L_i
\left( \begin{array}{cc} 0 & 0
\\ 0 & U_2 \end{array} \right) \right]_{N+1} = \left(
\begin{array}{c|c} 0 & 0 \\ \hline 0 & (L_i)_N \end{array}
\right) \label{a3} \eeq

and that the component $ U_2 $ is practically decoupled from the
others.

We must show that

\beq \left[ \left( \begin{array}{cc} U_1 & 0
\\ U_{12} & 0 \end{array} \right) L_i
\left( \begin{array}{cc} U_1^{\dagger} & U_{12}^{\dagger}
\\ 0 & 0 \end{array} \right) \right]_{N+1} = \left(
\begin{array}{c|c} (L_i)_{N+2} & 0 \\ \hline 0 & 0 \end{array}
\right) \label{a4} \eeq

We start considering the diagonal matrix $L_3$ :

\beq L_3 = \sum_{n_1, n_2} ( n_1 - n_2 ) | n_1, n_2 >< n_1, n_2 |
\label{a5} \eeq

The term

\beq U_1 L_3 U^{\dagger}_1 = \sum_{n_1,n_2} ( n_1 + 1 -n_2 ) | n_1,
n_2 >< n_1, n_2 | \label{a6} \eeq

once that it is restricted to the component with total number $ n_1
+ n_2 = N + 1 $ gives rise to

\beq ( U_1 L_3 U^{\dagger}_1 )_{N+1} = \sum_{k=0}^{N+1} ( N+2 - 2k)
| N+1-k, k >< N+1-k, k | \label{a7} \eeq

It is a diagonal matrix, having as entries $ ( N+2 ), .... , - N $.
To complete the representation of $ {( L_3 )}_{N+2} $, it is enough
to add a matrix element with value $ - ( N+ 2 ) $.

We note that, by construction, the off-diagonal elements are null

\beq U_1 L_3 U_{12}^{\dagger} = U_{12} L_3 U_1^{\dagger} = 0
\label{a8} \eeq

and therefore the remaining matrix element must arise from the term

\beq U_{12} L_3 U_{12}^{\dagger} = - \sum_{n_1 = n_2} ( n_2 + 1) |
n_1, 0 >< n_1, 0 | \label{a9} \eeq

When projecting the operator to the component $ n_1 = n_2 = ( N+1 )
$, we obtain :

\beq ( U_{12} L_3 U_{12}^{\dagger} )_{N+1} = - ( N+2 ) | N+1, 0 ><
N+1, 0 | \label{a10} \eeq

that is exactly the right term in the right position to complete the
representation

\beq \left( \begin{array}{c|c} ( L_3 )_{N+2} & 0 \\ \hline 0 & 0
\end{array} \right) \label{a11} \eeq

Now we are going to consider the case of $ L_{+} $

\beq L_{+} = a^{\dagger}_0 a_1 = \sum_{n_1, n_2} \sqrt{ ( n_1+1 )(
n_2+1 ) } | n_1+1, n_2 >< n_1, n_2+1 | \label{a12} \eeq

By construction we obtain

\begin{eqnarray}
 U_1 L_{+} U_1^{\dagger} &  = & \sum_{n_1, n_2} \sqrt{ ( n_1+2 )(
n_2+1 ) } | n_1+1, n_2 >< n_1, n_2+1 | \nonumber \\
( U_1 L_{+} U_1^{\dagger})_{N+1} &  = & \sum_{k=0}^{N} \sqrt{ (
N-k+2 ) ( k+1 ) } | N-k+1, k >< N-k, k+1 | \label{a13}
\end{eqnarray}

This term is part of the representation $ {( L_{+} )}_{N+2} $,
without the term $ k = N+1 $.

By looking at the explicit matrix $ L_{+} $, we expect that the term
$ k = N+1 $ must appear in the combination $ U_1 L_{+}
U_{12}^{\dagger} $. in fact it is easy to show that

\beq U_{12} L_{+} U_1^{\dagger} = U_{12} L_{+} U^{\dagger}_{12} = 0
\label{a14} \eeq

The explicit calculation gives for $ U_1 L_{+} U_{12}^{\dagger} $

\beq U_1 L_{+} U_{12}^{\dagger} = \sum_{n_1=n_2} \sqrt{ n_2+1 } | 0,
n_2 >< n_1, 0 | \label{a15} \eeq

and its projection to the term $ n_1 = n_2 = N+1 $ implies that

\beq ( U_1 L_{+} U_{12}^{\dagger} )_{N+1} = \sqrt{ N+2 } | 0, N+1 ><
N+1, 0 |  \label{a16} \eeq

that is what we need to complete the representation $ {( L_{+}
)}_{N+2} $, Analogous proof holds for $ L_{-}  $.

\end{document}